\documentclass{Interspeech}

\interspeechcameraready

\title{An Age-Agnostic System for Robust Speaker Verification}

\author[affiliation={1}]{Jiusi}{Zheng}
\author[affiliation={1}]{Vishwas}{Shetty}
\author[affiliation={1}]{Natarajan}{Balaji Shankar}
\author[affiliation={1}]{Abeer}{Alwan}

\affiliation{Dept of Electrical and Computer Engineering}{University of California, Los Angeles}{USA}
\email{{zheng94,shettyvishwas,balaji1312}@ucla.edu, alwan@ee.ucla.edu}

\usepackage{comment}
\usepackage{subcaption}
\usepackage{multirow}

\usepackage[table]{xcolor}
\usepackage{colortbl}
\definecolor{rowgray}{gray}{0.95}
\newcounter{rownumber} 
\newcommand{\rownumber}{\stepcounter{rownumber}\arabic{rownumber}} %

\begin{document}

\maketitle

\begin{abstract}
    In speaker verification (SV), the acoustic mismatch between children’s and adults' speech leads to suboptimal performance when adult-trained SV systems are applied to children's speaker verification (C-SV). While domain adaptation techniques can enhance performance on C-SV tasks, they often do so at the expense of significant degradation in performance on adults' SV (A-SV) tasks. In this study, we propose an Age Agnostic Speaker Verification (AASV) system that achieves robust performance across both C-SV and A-SV tasks. Our approach employs a domain classifier to disentangle age-related attributes from speech and subsequently expands the embedding space using the extracted domain information, forming a unified speaker representation that is robust and highly discriminative across age groups. Experiments on the OGI and VoxCeleb datasets demonstrate the effectiveness of our approach in bridging SV performance disparities, laying the foundation for inclusive and age-adaptive SV systems.
\end{abstract}

\section{Introduction}

Inclusiveness is important in speaker verification (SV) systems. Research shows that with the advancement of speech technology, 
94\% of children in the United States have used one or more voice assistants such as Siri and Google Assistant \cite{yarosh2018children,hoy2018alexa}. Beyond their traditional applications in security, SV systems hold great potential for enhancing children's interactions with technology, enabling seamless and user-friendly personalization. However, due to the acoustic mismatch between children’s and adults’ speech 
\cite{katz2001identification,ghai2010exploring,russell2007challenges}, state-of-the-art SV systems \cite{desplanques2020ecapa,zhou2021resnext,snyder2018x} which perform exceptionally well on benchmark adults' speech datasets often struggle with children's speech. To meet the diverse needs of users across different age groups, it is crucial to develop an inclusive SV system that delivers robust performance in both adults' SV (A-SV) and children's SV (C-SV) tasks, addressing the challenges posed by age-related acoustic variability.

To address the domain mismatch problem arising from the acoustic differences between children's and adults' speech, various approaches have been proposed in the context of C-SV \cite{shahnawazuddin2020domain, singh2024childaugment, aziz2023effective}. For instance, ChildAugment \cite{singh2024childaugment} mitigates this mismatch by transforming adults' speech to acoustically resemble children's speech. This is achieved by modifying vocal tract parameters such as formant frequencies and bandwidths, thereby enhancing model adaptation in zero-resource settings. Another prominent data augmentation technique involves voice conversion using a cycle-consistent GAN \cite{kaneko2017parallel}, which modifies adults' speech to mimic children's spectral and prosodic characteristics, effectively reducing the acoustic mismatch \cite{shahnawazuddin2021children}. 

In addition to data augmentation methods \cite{singh2024childaugment,shahnawazuddin2021children,shahnawazuddin2017effect, prasanna2010fast}, other domain adaptation methods have also been explored \cite{zhang2024speaker,shetty2025enhancing,li2022coral++,lin2023robust}. For example, CORAL++ extends CORAL by integrating deep neural networks to learn nonlinear transformations, effectively reducing the distribution mismatch between the source and target domain speech data \cite{li2022coral++}. However, a common limitation of these approaches is the significant performance trade-off between domains, as most of these methods compromise performance in the source domain while optimizing for the target domain. This issue is particularly pronounced in age-based domain mismatches, where the performance degradation in SV systems is more severe than in other cross-domain settings, such as cross-device \cite{huang2024robust}, cross-distance \cite{huang2024robust}, and cross-dialect \cite{huang2024robust,lin2023robust,thienpondt2020cross} scenarios. 

In this paper, we propose an Age Agnostic Speaker Verification (AASV) system that achieves strong generalization across both C-SV and A-SV tasks. The AASV system dynamically expands the embedding space and integrates complementary information from both adults and children SV models. Given the strong correlation between speaker embeddings and age \cite{koluguri2020meta}, we first employ ECAPA-TDNN \cite{desplanques2020ecapa} as the feature encoder to suppress linguistic information and obtain high-level speaker embeddings. A domain classifier is then optimized to disentangle age-related information and infer whether the speaker belongs to the adults or children domain, providing domain information for integration into the speaker embedding space. This information is subsequently employed to merge speaker representations, yielding an age agnostic speaker representation that is both robust and highly discriminative. To the best of our knowledge, despite existing research in the cross-domain SV \cite{huang2024robust,lin2023robust}, our work is the first to simultaneously report results for both C-SV and A-SV, establishing a new benchmark in the field.

\section{System Overview}

\begin{figure}[t]
  \centering
  \includegraphics[width=\linewidth]{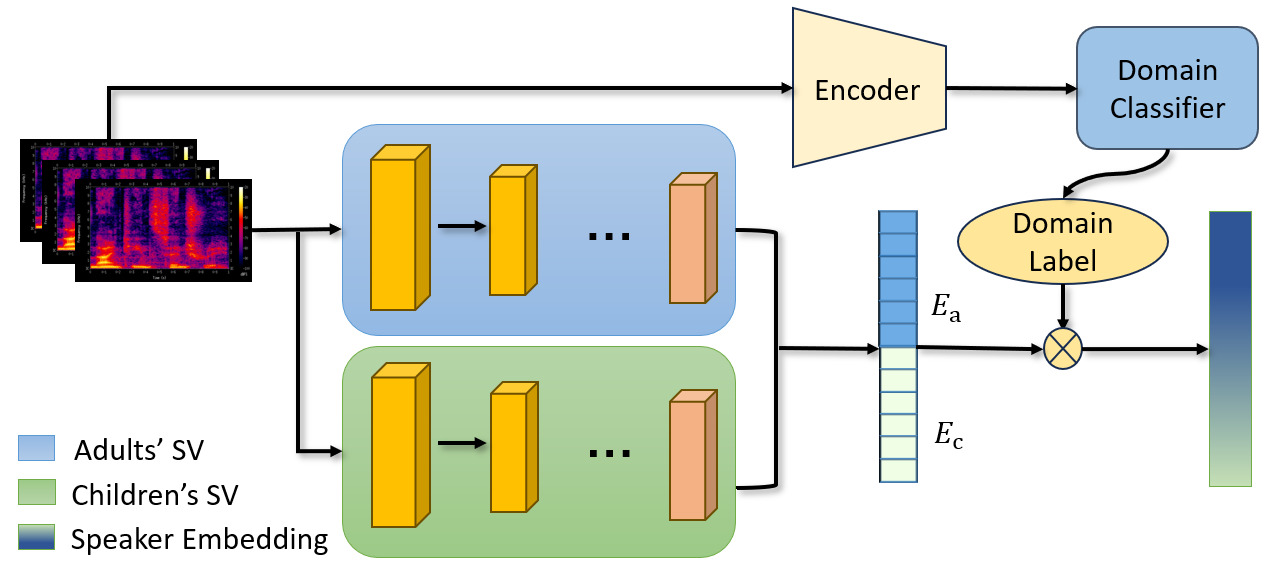}
  \caption{Workflow of the Age Agnostic Speaker Verification (AASV) system. Speaker embeddings $E_a$ and $E_c$ from A-SV and C-SV subsystems are fused with age-domain information from the domain classifier to generate robust speaker embeddings.
}
  \label{fig:AASV_workflow}
\end{figure}

\subsection{Domain-Specific Speaker Embedding Network}
We consider a neural network trained on adults' speech to map input speech signals to speaker embeddings for speaker verification (SV) tasks. This process optimizes a speaker embedding network, parameterized by \( \theta_{\text{a}} \), and a classification head with parameters \( \phi_{\text{a}} \). To enhance inter-class separability in the speaker embedding space, Additive Angular Margin loss (AAM-softmax) is employed to increase the angular margin between different speaker classes during the optimization of both components. The resulting speaker embedding network with parameters \( \theta_{\text{a}} \) specializes in extracting domain-specific speaker embeddings \( E_{\text{a}} \) tailored to adults' speech.

To adapt the network for children’s speech, we initiate a fine-tuning phase, initializing the network with the optimized parameters \( \theta_{\text{a}} \) and adapting it to child speech yielding updated parameters \( \theta_{\text{c}} \). A new classification head \( \phi_{\text{c}} \) is introduced and optimized to align with the adapted feature space. Rather than training a single cross-domain speaker embedding network on a mixed dataset, maintaining separate networks for different domains offers two key advantages: (1) Feature-level adaptation is more efficient than data-level mixing, as it facilitates knowledge transfer across domains without requiring the relearning of shared representations, and (2) In low-resource scenarios like C-SV, data imbalance caused by domain mixing can bias the model towards adults' speech, thereby undermining the performance on C-SV tasks.

\subsection{Age Representation Disentanglement}
\label{OGI_Domain_Classifier}
\begin{figure}[h]
    \centering
    \begin{subfigure}{0.23\textwidth}
        \centering
        \includegraphics[width=\linewidth]{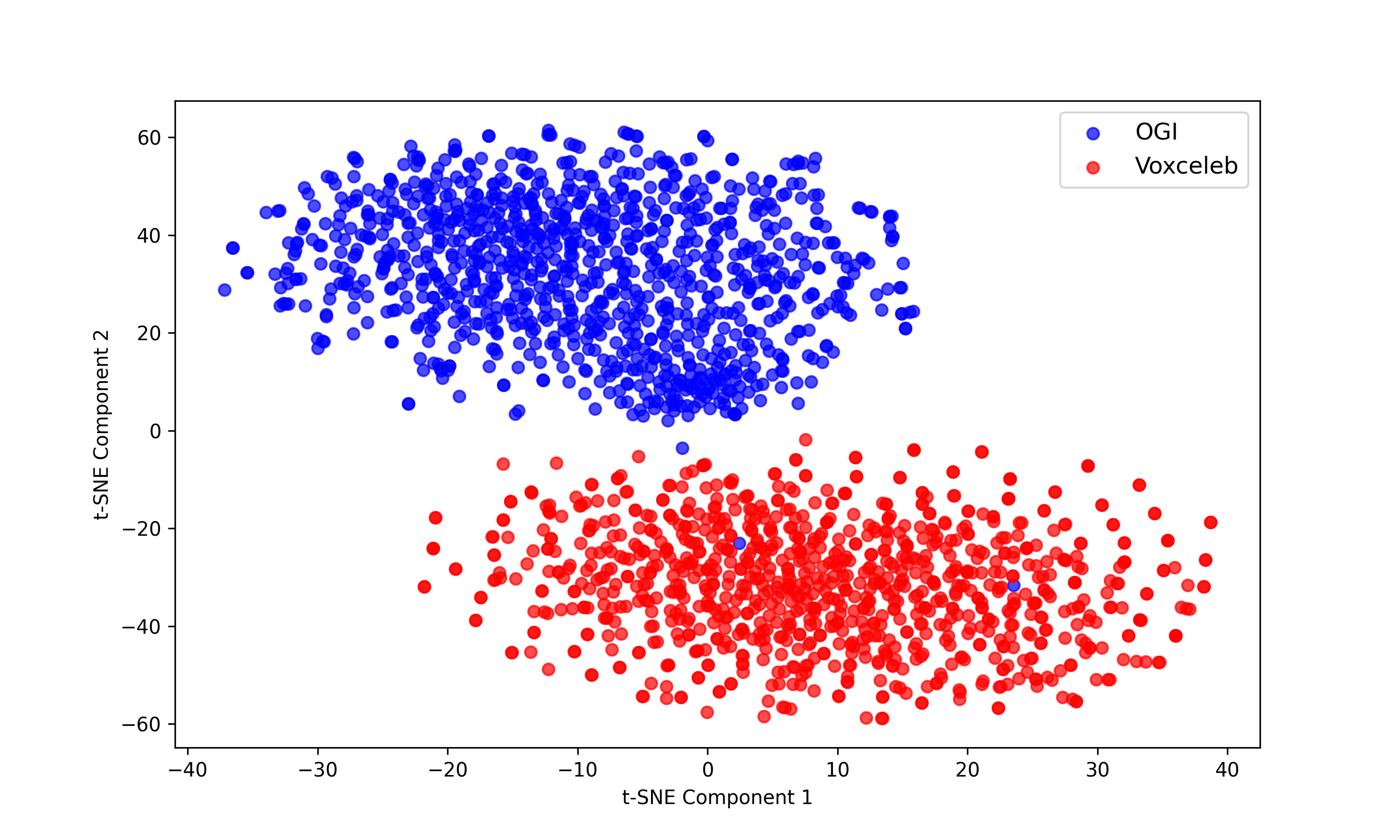}
        \caption{}
        \label{fig:sub1}
    \end{subfigure}
    \hfill
    \begin{subfigure}{0.23\textwidth}
        \centering
        \includegraphics[width=\linewidth]{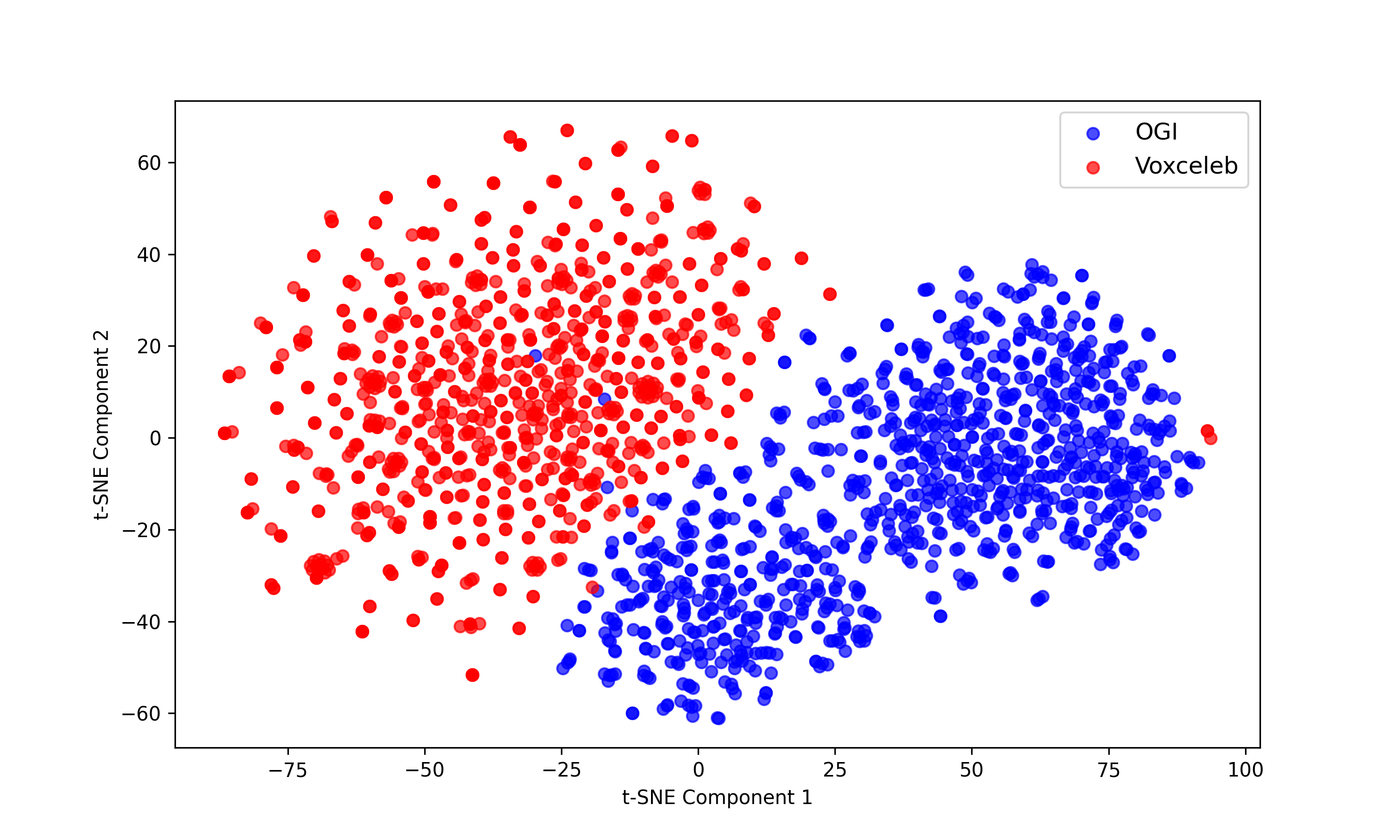}
        \caption{}
        \label{fig:sub2}
    \end{subfigure}
    
    \caption{t-SNE visualization of speaker embeddings. Blue and red points denote children's speech from OGI and adult speech from VoxCeleb, respectively. (a) t-SNE plot of younger children's group (K00-K06) vs. adults' speech. (b) t-SNE plot of older children's group (K07-K10) vs. adults' speech. The clear separation between domains highlights that speaker embeddings inherently capture age-related information.}
    \label{fig:tSNE}
\end{figure}

To extract age-related information from speech, as shown in Figure \ref{fig:AASV_workflow}, we design a domain classifier to learn domain labels. Considering the linguistic differences between children's and adults' speech datasets, the domain classifier may struggle to learn age-related speaker characteristics, potentially leading to misclassification based on content-related variations. To mitigate this issue, we used the ECAPA-TDNN\footnote{https://huggingface.co/speechbrain/spkrec-ecapa-voxceleb} model 
pre-trained on SV tasks as an encoder to transform speech samples of varying durations into fixed-length speaker embeddings. Figure \ref{fig:tSNE} presents t-SNE visualizations of the extracted embeddings, where children's and adults' speech are treated as distinct domains. Specifically, Figure \ref{fig:sub1} illustrates the clustering results when defining younger children's speech (K00-K06 age groups in the OGI dataset \cite{shobaki2000ogi}) as a separate domain, with adults' speech drawn from VoxCeleb2 \cite{chung2018voxceleb2}. In contrast, Figure \ref{fig:sub2} illustrates the distribution when older children's speech (K07-K10 age groups in the OGI dataset) is considered instead. The results indicate that embeddings from younger children's speech exhibit greater divergence from embeddings drawn from adults' speech, forming two well-separated and compact clusters in Figure \ref{fig:sub1}. Leveraging this divergence, we train the domain classifier using speech samples from younger children and adults, thereby enhancing its ability to distinguish between domains effectively.

\subsection{Multi-space Embedding Integration}

We employ a multi-space embedding integration approach that ensures the formation of robust speaker embeddings. Specifically, we extract two sets of speaker embeddings:  \(\mathbf{E}_a \in \mathbb{R}^d\),  obtained from the speaker embedding network trained on adults' speech, and \(\mathbf{E}_c \in \mathbb{R}^d\) from the speaker embedding network fine-tuned on children's speech, where \( d \) represents the dimensionality of the speaker embeddings. To integrate these embeddings, we leverage the softmax output of the domain classifier, which provides a probabilistic estimation of domain affiliation. Let the domain classifier output be $\mathbf{p} = [p_c, p_a]$, where $\textbf{p} \in R^2$, and, $p_c$ and $p_a$ are scalars representing the probability the input audio belongs to the children's and adults' domains, respectively. These domain scores act as dynamic fusion coefficients, guiding the weighted combination of the two embeddings within the expanded embedding space. We concatenate the weighted embeddings along the feature dimension, expanding the embedding space and forming a robust representation:

\begin{equation}
\mathbf{E}_{\text{R}} = \left[ p_c \mathbf{E}_c ; p_a \mathbf{E}_a \right] \in \mathbb{R}^{2d}.
\end{equation}

The resulting representation $\mathbf{E}_{\text{R}}$ captures speaker characteristics across both age domains. When the speakers age domain is uncertain, the domain classifier balances the contributions of both embedding networks, mitigating errors caused by incorrect domain classification.
  
\section{Experimental Setup}
\subsection{Datasets}

We use the VoxCeleb dataset as the adults' speech dataset \cite{nagrani2017voxceleb,chung2018voxceleb2}. VoxCeleb is a large-scale corpus of speech collected from YouTube, comprising two subsets, VoxCeleb1 and VoxCeleb2, and contains audio samples collected from a range of acoustic conditions. The OGI dataset \cite{shobaki2000ogi} is used as the children's speech dataset. The OGI corpus includes both spontaneous and scripted speech data from approximately 1,100 children, spanning ages from kindergarten (\textit{K00}) to grade 10 (\textit{K10}). Each age group contains about 100 speakers. In our experiments, we utilize only the scripted speech portion of the OGI dataset.

We employ different train-eval splits across various datasets for our experiments. For A-SV tasks, models pre-trained on the VoxCeleb1 training set and the VoxCeleb2 development set are evaluated solely on the VoxCeleb1-O (Original) test set. Models trained on the VoxCeleb2 set are evaluated on the VoxCeleb1-O, VoxCeleb1-E (Extended), and VoxCeleb1-H (Hard) test sets for a more comprehensive assessment. For C-SV tasks, we utilize the train-eval splits of the OGI dataset in \cite{fan2024benchmarking}. The training set, drawn from grades K00 to K10 age groups, comprises data from 866 speakers, totaling approximately 24 hours of speech. Additionally, the domain classifier is trained using a subset of the OGI training set, specifically grades K00 to K06 as discussed in Section \ref{OGI_Domain_Classifier}, alongside the VoxCeleb2 development set. The OGI test set consists of 4,000 evaluation trials per age group, with 2,000 positive and 2,000 negative trials each, resulting in 44,000 SV trials from 225 speakers.

\subsection{Speaker Verification Systems}
We employ two configurations of the ECAPA-TDNN \cite{desplanques2020ecapa} model for SV tasks, with convolutional frame layers using 512 or 1024 channels to provide a comprehensive evaluation. All experiments are conducted in the SpeechBrain framework \cite{speechbrain}. 
The first ECAPA-TDNN model uses 512 channels per convolutional frame layer, resulting in approximately 6.2M parameters. This model was pre-trained on VoxCeleb2 development set and is referred to as \textit{A-SV-Small}. Then we fine-tune \textit{A-SV-Small} on the OGI dataset to obtain \textit{C-SV-Small}. As the smaller ECAPA-TDNN is only pre-trained on VoxCeleb2, we report results on VoxCeleb1-O, VoxCeleb1-H, and VoxCeleb1-E.

We also employ a second ECAPA-TDNN configuration with 1024 channels per convolutional frame layer, resulting in approximately 20.8M parameters. This model was pre-trained on the VoxCeleb1 training set and VoxCeleb2 development set, and we refer to it as \textit{A-SV-Large}. This model is fine-tuned on the OGI dataset to obtain \textit{C-SV-Large}. Since \textit{A-SV-Large} was pre-trained on both VoxCeleb1 and VoxCeleb2, we only use VoxCeleb1-O as the test set for A-SV to ensure data separation.

\begin{table}[t!]
\centering
\caption{\textit{Accuracy of domain classifiers trained on mixed datasets with varying proportions of children’s and adults’ speech. The model is evaluated on separate children’s and adults’ test sets, simulating the domain classifier’s application in SV tasks. The children’s test set is a mixture of K00-K10 OGI test sets and the adults’ test set is VoxCeleb1-O.}}
\resizebox{0.8\columnwidth}{!}{%
\renewcommand{\arraystretch}{1.1}
\begin{tabular}{lcccc}
\hline 
\multirow{2}{*}{} & \multicolumn{2}{c}{\# Utterances} & \multicolumn{2}{c}{Accuracy} \\ \cline{2-3} \cline{4-5} 
                  & Adults  & Children  & Adults  & Children  \\ \hline
\textit{1:1}   & 20,000   & 20,000   & 42.1\%      & 100.0\%     \\
\textit{2:1}   & 40,000   & 20,000   & 70.6\%      & 99.9\%     \\ 
\textit{3:1}   & 60,000   & 20,000   & 87.5\%      & 99.9\%     \\ 
\textit{4:1}   & 80,000   & 20,000   & 89.7\%      & 99.7\%     \\ 
\textit{5:1}   & 100,000   & 20,000   & 95.0\%      & 99.6\%     \\ \hline
\end{tabular}%
}
\label{tab:doamin_classifier_outcome}
\end{table}

\subsection{Training Details and Evaluation Metric}
During fine-tuning of the pretrained ECAPA-TDNN model on the OGI dataset, which includes speakers from the K00 to K10 age groups, the following configuration is employed: The network is optimized using the Adam optimizer with a weight decay of $2 \times 10^{-6}$. The learning rate is set to 0.001, with a cyclic annealing schedule, where the base learning rate is set to $1 \times 10^{-8}$ and the maximum learning rate is 0.001. Fine-tuning is conducted for 15 epochs, with a batch size of 16. The input features consisted of 80-dimensional filter bank coefficients, computed every 10 ms with a 25 ms window. The audio duration for each utterance is 2 seconds in the training set. To improve the robustness, we applied four data augmentation techniques: additive noise, reverberation with room impulse responses (RIR), frequency masking, and time masking.

\begin{table*}[t!]
    \centering
    \caption{\textit{EER (\%) comparison of verification systems. K00–K10 represent increasing age groups. A-SV and C-SV denote Adults' and Children's verification systems, respectively, while AASV is our proposed Age Agnostic Speaker Verification system. WSE refers to the Weight Space Ensemble (WSE) model. Systems labeled "-Large" (20.8M parameters) were trained on VoxCeleb1 and are not evaluated on VoxCeleb1-E and VoxCeleb1-H (indicated by ‘-’). Systems labeled "-Small" (6.2M parameters) are trained on VoxCeleb2. DC denotes the domain classifier, PreEmbed represents pre-trained speaker embeddings. All C-SV systems used the same OGI training set. Rows 4 and 9 show that our methods maintain balanced performance on both child and adult test sets.
}}
    \label{tab:performance}
    \resizebox{\textwidth}{!}{ 
    \renewcommand{\arraystretch}{1.1}
    
    \begin{tabular}{clccccccccccccccc}
        \toprule
        \addlinespace[0.5em]
        \multirow{2}{*}{\textbf{\#}} &\multirow{2}{*}{\textbf{Systems}} & \multicolumn{11}{c}{\textbf{OGI (Children's Speech)}} & \multicolumn{3}{c}{\textbf{VoxCeleb (Adults' Speech)}} \\
        \cmidrule(lr){3-13} \cmidrule(lr){14-16}
        & & K00 & K01 & K02 & K03 & K04 & K05 & K06 & K07 & K08 & K09 & K10 & VoxCeleb1-O & VoxCeleb1-E & VoxCeleb1-H \\
        \midrule
        \rownumber & A-SV-Large & 30.65 & 30.55 & 31.75 & 26.85 & 23.90 & 12.55 & 13.10 & 9.15 & 6.15 & 5.80 & 6.15 & 0.80 & - & - \\
        \rowcolor{gray!20} \rownumber & C-SV-Large & 8.90 & 10.10 & 8.55 & 6.40 & 6.45 & 3.60 & 3.20 & 2.55 & 2.40 & 1.60 & 3.20 & 7.01 & - & - \\
        \rownumber & ChildAugment \cite{singh2024childaugment} & 29.30 & 29.65 & 29.95 & 24.55 & 23.55 & 12.30 & 13.15 & 8.9 & 6.10 & 5.65 & 5.75 & 1.19 & 1.49 & 2.98 \\
        \rowcolor{gray!20} \rownumber & \textbf{AASV-Large (Ours}) & 8.90 & 10.20 & 8.55 & 6.40 & 6.45 & 3.60 & 3.20 & 2.55 & 2.30 & 1.70 & 4.00 & 1.89 & - & - \\ 
        \bottomrule
        \rownumber & A-SV-Small & 31.95 & 34.40 & 32.30 & 29.85 & 28.80 & 15.40 & 14.90 & 11.60 & 10.10 & 9.15 & 9.60 & 1.19 & 1.31 & 2.48 \\
        \rowcolor{gray!20} \rownumber & C-SV-Small & 9.50 & 9.50 & 8.45 & 6.60 & 5.60 & 3.80 & 3.30 & 2.80 & 2.50 & 1.80 & 3.00 & 10.21 & 9.55 & 13.38 \\
        \rownumber & Stable Learning \cite{zhang2024speaker} & 8.55 & 9.6 & 8.4 & 6.55 & 6.35 & 4.00 & 3.05 & 2.8 & 2.15 & 1.6 & 3.15 & 10.26 & 9.55 & 13.38 \\
        \rowcolor{gray!20} \rownumber & WSE \cite{lin2023robust} & 17.25 & 18.40 & 16.05 & 14.30 & 13.95 & 7.25 & 5.80 & 4.35 & 3.15 & 3.55 & 3.15 & 1.92 & 2.13 & 3.44 \\
        \rownumber & \textbf{AASV-Small (Ours}) & 9.50 & 9.60 & 8.45 & 6.65 & 5.55 & 3.80 & 3.30 & 2.80 & 2.40 & 1.90 & 3.80 & 2.25 & - & - \\ 
        \rowcolor{gray!20} & \quad w/o DC & 16.65 & 17.60 & 16.35 & 13.80 & 13.95 & 8.00 & 5.90 & 4.75 & 4.45 & 3.30 & 4.75 & 1.85 & 2.13 & 3.63 \\
        & \quad w/o DC + PreEmbed & 19.80 & 22.15 & 21.10 & 19.45 & 19.65 & 10.20 & 9.00 & 6.00 & 5.75 & 4.90 & 5.10 & 2.43 & 2.66 & 4.56 \\
        \rowcolor{gray!20} \rownumber & AASV-Small + PreEmbed &11.7 & 14.45 & 12.54 & 11.00 & 11.45 & 8.40 & 7.20 & 5.00 & 4.35 & 3.95 & 5.15 & 1.95 & - & - \\ 
        \bottomrule
        \addlinespace[0.5em]
    \end{tabular}
    }
\end{table*}

For the A-SV systems, we directly utilize the open-source ECAPA-TDNN models pre-trained on the Voxceleb dataset, AASV-Large\footnote{https://huggingface.co/speechbrain/spkrec-ecapa-voxceleb} and AASV-Small\footnote{https://huggingface.co/yangwang825/ecapa-tdnn-vox2}. The domain classifier is trained using the cross-entropy loss function, achieving 99\% overall accuracy and an F1 score of 0.99. Additionally, data augmentation is also applied during the training of the domain classifier, where one of the four augmentations mentioned earlier is randomly selected per iteration. Equal Error Rate (EER) is used to evaluate the performance of the SV systems.

\section{Results and Discussion}
\subsection{Influence of the domain classifier}
\label{Data Bias}
Table~\ref{tab:doamin_classifier_outcome} presents results from training the domain classifier with varying amounts of adult speech, while keeping children's data fixed at 20,000 utterances. As the adult-to-child ratio increases from 1:1 to 5:1, accuracy on the adult test set rises from 42.1\% to 95.0\%, with minimal change on the children’s set. No further improvement is observed beyond a 5:1 ratio. The need for a higher proportion of adults' speech data in the training set is likely due to the larger size and greater diversity of the VoxCeleb dataset compared to the OGI dataset. VoxCeleb includes a broader range of accents, speaker ages, and acoustic conditions, which likely contributes to the need for a larger sample to ensure effective domain classification.

Since the domain classifier was initialized with the ECAPA-TDNN encoder pre-trained on VoxCeleb1 and VoxCeleb2 (excluding VoxCeleb1-O), we did not report the results of the AASV-Large and AASV-Small systems on VoxCeleb1-E and VoxCeleb-H to avoid potential bias. Additionally, we noticed that although misclassification occurred with the domain classifier, their impact on the EER in SV tasks was smaller than the misclassification rate itself. Specifically, although the domain classifier achieved 95\% accuracy on VoxCeleb1, the remaining 5\% error rate led to only a 1.09\% absolute increase in EER for the AASV-Large system compared to the A-SV-Large system and a 1.06\% absolute increase for the AASV-Small system compared to the A-SV-Small system as shown in Table~\ref{tab:performance}. This is likely because some cross-speaker test pairs were misclassified into different domains, which amplified embedding differences and reduced cosine similarity, without causing verification errors.

\subsection{Performance on Children's and Adults' Speech}
\subsubsection{The performance of A-SV and C-SV systems}
Table~\ref{tab:performance} shows the EER performance of different SV systems across various test sets. The \textit{A-SV-Large} system (row 1) exhibits a general decrease in EER as the age of speakers in the test sets increases, with EER dropping from 30.65\% on the K00 test set to 6.15\% on the K10 set, and further to 0.8\% on the VoxCeleb1-O test set. In contrast, the \textit{C-SV-Large} system (row 2) achieves lower EERs on the OGI year-wise test sets, decreasing from 8.9\% on K00 to 3.2\% on K10; however, its EER on the VoxCeleb1-O test set increases to 7.01\%, reflecting an absolute increase of 6.21\% compared to \textit{A-SV-Large}. A similar trend is observed when comparing \textit{A-SV-Small} and \textit{C-SV-Small} systems (rows 5 and 6). Overall, A-SV systems perform well on the VoxCeleb test sets but struggle on the OGI year-wise test sets due to domain mismatch. Conversely, C-SV systems significantly reduce EER on the OGI year-wise test sets; but this improvement comes at the cost of catastrophic forgetting of the source-domain knowledge.

\subsubsection{The performance of the proposed AASV systems}
\begin{figure}[t]
  \centering
  \includegraphics[width=\linewidth]{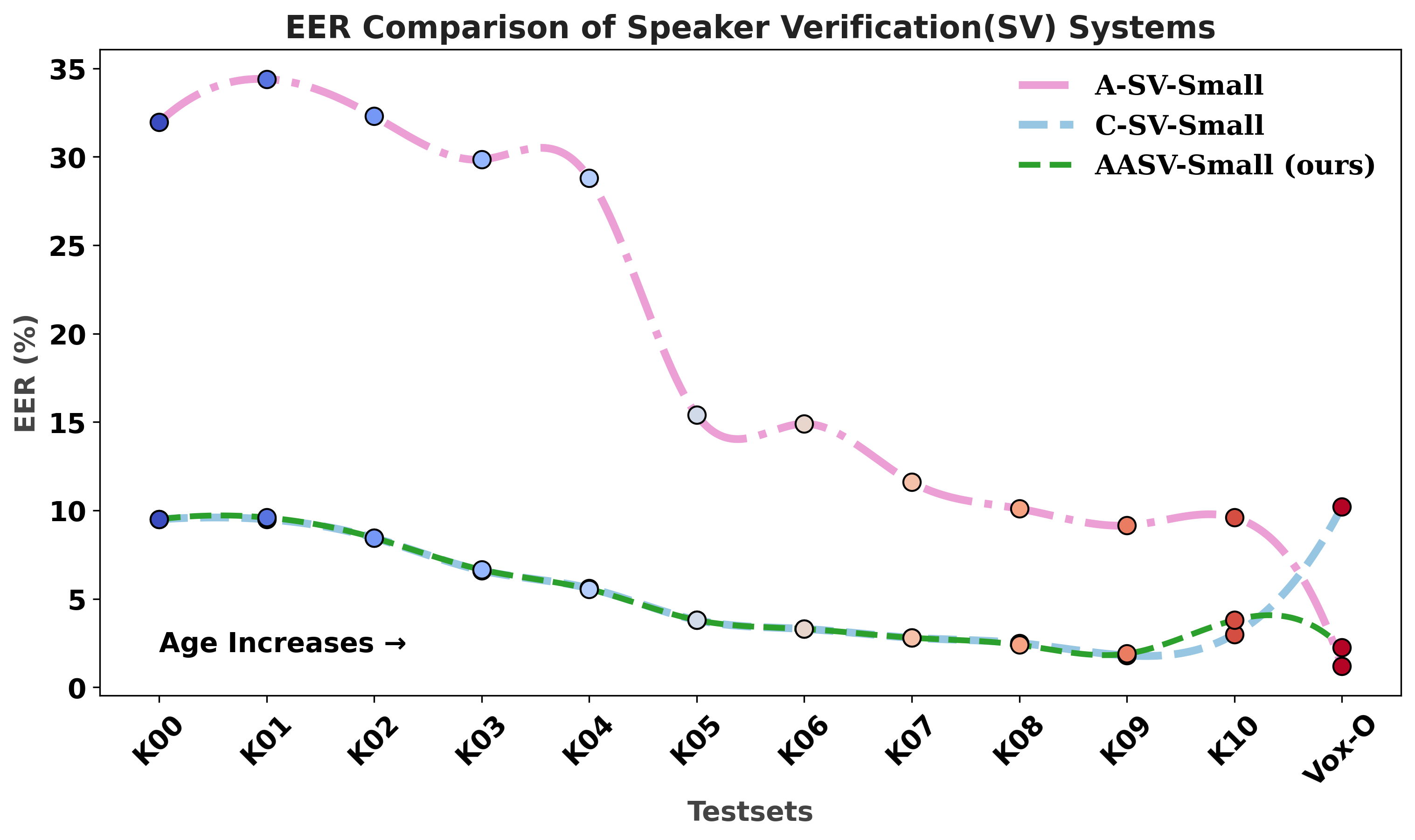}
  \caption{Performance of the A-SV-Small, C-SV-Small, and proposed AASV-Small systems across year-wise OGI test sets and the VoxCeleb1 test set.}
  \label{fig:EER_Graph}
\end{figure}

The proposed \textit{AASV-Large} and \textit{AASV-Small} systems achieve competitive performance across both children’s and adults’ test sets. As illustrated in Figure~\ref{fig:EER_Graph}, the absolute increase in EER for the \textit{AASV-Small} system compared to the \textit{C-SV-Small} system across the K00-K09 age groups is within 0.1\%. For the K10 test set, the absolute increase in EER is 0.8\%. Meanwhile, the \textit{AASV-Small} system exhibits an absolute increase of 1.06\% in EER compared to the \textit{A-SV-Small} system on the Voxceleb1-O test set, substantially lower than the 9.02\% absolute rise in EER observed for the \textit{C-SV-Small} system on the Voxceleb1-O test set. We also compare our approach with three existing methods: ChildAugment~\cite{singh2024childaugment}, Stable Learning~\cite{zhang2024speaker}, and Weight Space Ensemble (WSE)~\cite{lin2023robust}. As shown in Table~\ref{tab:performance}, out-of-domain data augmentation method ChildAugment yields limited improvements on C-SV test sets while maintaining stable performance on A-SV test sets. In contrast, Stable Learning method improves C-SV but degrades A-SV performance significantly. Rows 8 and 9 of Table 2 compare our approach with WSE, AASV-Small system outperforms the WSE-based model on 10 out of 11 OGI year-wise test sets, demonstrating better generalization across child age groups. The WSE method achieves slightly better results on the K10 and VoxCeleb1-O test sets, with EERs of 3.5\% and 1.92\%, respectively.

\subsection{ Ablation study on embedding integration strategies}
To demonstrate the importance of the domain classifier in the AASV systems, we present the EER data without the domain classifier (\textit{w/o DC}) in the Table~\ref{tab:performance}. We note that directly integrating the speaker embeddings of the C-SV and A-SV subsystems without the domain classifier results in a significant performance drop on the OGI year-wise test sets and a slight performance improvement on Voxceleb1-O testset. Inspired by the findings in \cite{sari2019pre} regarding the utility of pre-trained speaker embeddings in providing useful information across tasks different from pre-training, as well as their robustness when applied to target-domain tasks, we also explored incorporating speaker embeddings from the adults’ speech pre-trained SV model during C-SV model fine-tuning (\textit{+~PreEmbed} in Table~\ref{tab:performance}), but this approach did not yield satisfactory results in our case.

\section{Conclusion}
In this paper, we propose the Age Agnostic Speaker Verification (AASV) system, which aims to enhance the inclusiveness of speaker verification systems by achieving competitive performance across both children's and adults' verification tasks. We first employ the domain classifier to obtain domain information, then integrate the speaker-related information extracted from both the children's and adults' speaker verification systems with age-related information. This fusion ultimately leads to the formation of a robust speaker embedding that is adaptable to both domains. Experimental results on the OGI and VoxCeleb datasets demonstrate that the AASV system achieves competitive results across different age groups. Future work will focus on more effective extraction of age features from speech and narrowing the performance gap between children's and adults' verification tasks.

\section{Acknowledgment}
The research is supported in part by the NSF and the IES, U.S. Department of Education (DoE), through Grant R305C240046 to the U. at Buffalo. The opinions expressed are those of the authors and do not represent views of the IES,  DoE, or the NSF.

\bibliographystyle{IEEEtran}
\bibliography{mybib}

\begin{thebibliography}{10}
\providecommand{\url}[1]{#1}
\csname url@samestyle\endcsname
\providecommand{\newblock}{\relax}
\providecommand{\bibinfo}[2]{#2}
\providecommand{\BIBentrySTDinterwordspacing}{\spaceskip=0pt\relax}
\providecommand{\BIBentryALTinterwordstretchfactor}{4}
\providecommand{\BIBentryALTinterwordspacing}{\spaceskip=\fontdimen2\font plus
\BIBentryALTinterwordstretchfactor\fontdimen3\font minus \fontdimen4\font\relax}
\providecommand{\BIBforeignlanguage}[2]{{%
\expandafter\ifx\csname l@#1\endcsname\relax
\typeout{** WARNING: IEEEtran.bst: No hyphenation pattern has been}%
\typeout{** loaded for the language `#1'. Using the pattern for}%
\typeout{** the default language instead.}%
\else
\language=\csname l@#1\endcsname
\fi
#2}}
\providecommand{\BIBdecl}{\relax}
\BIBdecl

\bibitem{yarosh2018children}
S.~Yarosh, S.~Thompson, K.~Watson, A.~Chase, A.~Senthilkumar, Y.~Yuan, and A.~B. Brush, ``Children asking questions: speech interface reformulations and personification preferences,'' in \emph{Proceedings of the 17th ACM conference on interaction design and children}, 2018, pp. 300--312.

\bibitem{hoy2018alexa}
M.~B. Hoy, ``Alexa, siri, cortana, and more: an introduction to voice assistants,'' \emph{Medical reference services quarterly}, vol.~37, no.~1, pp. 81--88, 2018.

\bibitem{katz2001identification}
W.~F. Katz and P.~F. Assmann, ``Identification of children's and adults' vowels: Intrinsic fundamental frequency, fundamental frequency dynamics, and presence of voicing,'' \emph{Journal of Phonetics}, vol.~29, no.~1, pp. 23--51, 2001.

\bibitem{ghai2010exploring}
S.~Ghai and R.~Sinha, ``Exploring the effect of differences in the acoustic correlates of adults' and children's speech in the context of automatic speech recognition,'' \emph{EURASIP Journal on Audio, Speech, and Music Processing}, vol. 2010, pp. 1--15, 2010.

\bibitem{russell2007challenges}
M.~J. Russell and S.~D'Arcy, ``Challenges for computer recognition of children2s speech.'' \emph{SLaTE}, vol. 108, p. 111, 2007.

\bibitem{desplanques2020ecapa}
B.~Desplanques, J.~Thienpondt, and K.~Demuynck, ``Ecapa-tdnn: Emphasized channel attention, propagation and aggregation in tdnn based speaker verification,'' in \emph{Interspeech 2020}, 2020, pp. 3830--3834.

\bibitem{zhou2021resnext}
T.~Zhou, Y.~Zhao, and J.~Wu, ``Resnext and res2net structures for speaker verification,'' in \emph{2021 IEEE Spoken Language Technology Workshop (SLT)}.\hskip 1em plus 0.5em minus 0.4em\relax IEEE, 2021, pp. 301--307.

\bibitem{snyder2018x}
D.~Snyder, D.~Garcia-Romero, G.~Sell, D.~Povey, and S.~Khudanpur, ``X-vectors: Robust dnn embeddings for speaker recognition,'' in \emph{2018 IEEE international conference on acoustics, speech and signal processing (ICASSP)}.\hskip 1em plus 0.5em minus 0.4em\relax IEEE, 2018, pp. 5329--5333.

\bibitem{shahnawazuddin2020domain}
S.~Shahnawazuddin, W.~Ahmad, N.~Adiga, and A.~Kumar, ``In-domain and out-of-domain data augmentation to improve children’s speaker verification system in limited data scenario,'' in \emph{ICASSP 2020-2020 IEEE International Conference on Acoustics, Speech and Signal Processing (ICASSP)}.\hskip 1em plus 0.5em minus 0.4em\relax IEEE, 2020, pp. 7554--7558.

\bibitem{singh2024childaugment}
V.~P. Singh, M.~Sahidullah, and T.~Kinnunen, ``Childaugment: Data augmentation methods for zero-resource children's speaker verification,'' \emph{The Journal of the Acoustical Society of America}, vol. 155, no.~3, pp. 2221--2232, 2024.

\bibitem{aziz2023effective}
S.~Aziz and S.~Shahnawazuddin, ``Effective preservation of higher-frequency contents in the context of short utterance based children’s speaker verification system,'' \emph{Applied Acoustics}, vol. 209, p. 109420, 2023.

\bibitem{kaneko2017parallel}
T.~Kaneko and H.~Kameoka, ``Parallel-data-free voice conversion using cycle-consistent adversarial networks,'' \emph{arXiv preprint arXiv:1711.11293}, 2017.

\bibitem{shahnawazuddin2021children}
S.~Shahnawazuddin, W.~Ahmad, N.~Adiga, and A.~Kumar, ``Children's speaker verification in low and zero resource conditions,'' \emph{Digital Signal Processing}, vol. 116, p. 103115, 2021.

\bibitem{shahnawazuddin2017effect}
S.~Shahnawazuddin, N.~Adiga, and H.~K. Kathania, ``Effect of prosody modification on children's asr,'' \emph{IEEE Signal Processing Letters}, vol.~24, no.~11, pp. 1749--1753, 2017.

\bibitem{prasanna2010fast}
S.~M. Prasanna, D.~Govind, K.~S. Rao, and B.~Yegnanarayana, ``Fast prosody modification using instants of significant excitation,'' in \emph{Speech Prosody 2010-Fifth International Conference}, 2010.

\bibitem{zhang2024speaker}
J.~Zhang, J.~Ma, X.~Guo, L.~Li, and L.~He, ``A speaker recognition method based on stable learning,'' in \emph{ICASSP 2024-2024 IEEE International Conference on Acoustics, Speech and Signal Processing (ICASSP)}.\hskip 1em plus 0.5em minus 0.4em\relax IEEE, 2024, pp. 10\,221--10\,225.

\bibitem{shetty2025enhancing}
V.~M. Shetty, J.~Zheng, S.~M. Lulich, and A.~Alwan, ``Enhancing age-related robustness in children speaker verification,'' in \emph{ICASSP 2025-2025 IEEE International Conference on Acoustics, Speech and Signal Processing (ICASSP)}.\hskip 1em plus 0.5em minus 0.4em\relax IEEE, 2025, pp. 1--5.

\bibitem{li2022coral++}
R.~Li, W.~Zhang, and D.~Chen, ``The coral++ algorithm for unsupervised domain adaptation of speaker recognition,'' in \emph{ICASSP 2022-2022 IEEE International Conference on Acoustics, Speech and Signal Processing (ICASSP)}.\hskip 1em plus 0.5em minus 0.4em\relax IEEE, 2022, pp. 7172--7176.

\bibitem{lin2023robust}
W.~Lin and M.-W. Mak, ``Robust speaker verification using deep weight space ensemble,'' \emph{IEEE/ACM Transactions on Audio, Speech, and Language Processing}, vol.~31, pp. 802--812, 2023.

\bibitem{huang2024robust}
W.~Huang, B.~Han, S.~Wang, Z.~Chen, and Y.~Qian, ``Robust cross-domain speaker verification with multi-level domain adapters,'' in \emph{ICASSP 2024-2024 IEEE International Conference on Acoustics, Speech and Signal Processing (ICASSP)}.\hskip 1em plus 0.5em minus 0.4em\relax IEEE, 2024, pp. 11\,781--11\,785.

\bibitem{thienpondt2020cross}
J.~Thienpondt, B.~Desplanques, and K.~Demuynck, ``Cross-lingual speaker verification with domain-balanced hard prototype mining and language-dependent score normalization,'' in \emph{Interspeech 2020}, 2020, pp. 756--760.

\bibitem{koluguri2020meta}
N.~R. Koluguri, M.~Kumar, S.~H. Kim, C.~Lord, and S.~Narayanan, ``Meta-learning for robust child-adult classification from speech,'' in \emph{ICASSP 2020-2020 IEEE International Conference on Acoustics, Speech and Signal Processing (ICASSP)}.\hskip 1em plus 0.5em minus 0.4em\relax IEEE, 2020, pp. 8094--8098.

\bibitem{shobaki2000ogi}
K.~Shobaki, J.-P. Hosom, and R.~Cole, ``The ogi kids’ speech corpus and recognizers,'' in \emph{Proc. of ICSLP}.\hskip 1em plus 0.5em minus 0.4em\relax Citeseer, 2000, pp. 564--567.

\bibitem{chung2018voxceleb2}
J.~S. Chung, A.~Nagrani, and A.~Zisserman, ``Voxceleb2: Deep speaker recognition,'' in \emph{Interspeech 2018}, 2018, pp. 1086--1090.

\bibitem{nagrani2017voxceleb}
A.~Nagrani, J.~S. Chung, and A.~Zisserman, ``Voxceleb: A large-scale speaker identification dataset,'' in \emph{Interspeech 2017}, 2017, pp. 2616--2620.

\bibitem{fan2024benchmarking}
R.~Fan, N.~{Balaji Shankar}, and A.~Alwan, ``Benchmarking children's asr with supervised and self-supervised speech foundation models,'' in \emph{Interspeech}, 2024, pp. 5173--5177.

\bibitem{speechbrain}
M.~Ravanelli, T.~Parcollet, P.~Plantinga, A.~Rouhe, S.~Cornell, L.~Lugosch, C.~Subakan, N.~Dawalatabad, A.~Heba, J.~Zhong, J.-C. Chou, S.-L. Yeh, S.-W. Fu, C.-F. Liao, E.~Rastorgueva, F.~Grondin, W.~Aris, H.~Na, Y.~Gao, R.~D. Mori, and Y.~Bengio, ``{SpeechBrain}: A general-purpose speech toolkit,'' 2021, arXiv:2106.04624.

\bibitem{sari2019pre}
L.~Sar{\i}, S.~Thomas, M.~Hasegawa-Johnson, and M.~Picheny, ``Pre-training of speaker embeddings for low-latency speaker change detection in broadcast news,'' in \emph{ICASSP 2019-2019 IEEE International Conference on Acoustics, Speech and Signal Processing (ICASSP)}.\hskip 1em plus 0.5em minus 0.4em\relax IEEE, 2019, pp. 6286--6290.

\end{thebibliography}

\end{document}